\documentclass[12pt,leqno]{article}
\usepackage{amsfonts}
\usepackage{amsmath, amsthm, amsfonts, amssymb, color}
\usepackage{mathrsfs}
\usepackage{color}
\theoremstyle{definition}

\newcommand{\scr}[1]{\mathscr #1}
\definecolor{wco}{rgb}{0.5,0.2,0.3}

\numberwithin{equation}{section} \theoremstyle{remark}

\newcommand{\ua}{\uparrow}

\title{{\bf Higher Order Eigenvalues for Non-Local Schr\"odinger Operators}\footnote{Supported in
 part by  NNSFC (11431014).} }
 \author{{\bf    Niels Jacob$^{b)}$, Feng-Yu Wang$^{a),b)}$  }\\
\footnotesize{ $^{a)}$ Center for Applied Mathematics, Tianjin University, Tianjin 300072, China }\\
\footnotesize{$^{b)}$ Department of Mathematics,
Swansea University, Singleton Park, SA2 8PP, United Kingdom}
\\
\footnotesize{  N.Jacob@swasnea.ac.uk, wangfy@tju.edu.cn, F.-Y.Wang@swansea.ac.uk}}
\begin{document}
\allowdisplaybreaks
\def\R{\mathbb R}  \def\ff{\frac} \def\ss{\sqrt} \def\B{\mathbf
B} \def\W{\mathbb W}
\def\N{\mathbb N} \def\kk{\kappa} \def\m{{\bf m}}
\def\ee{\varepsilon}\def\ddd{D^*}
\def\dd{\delta} \def\DD{\Delta} \def\vv{\varepsilon} \def\rr{\rho}
\def\<{\langle} \def\>{\rangle} \def\GG{\Gamma} \def\gg{\gamma}
  \def\nn{\nabla} \def\pp{\partial} \def\E{\mathbb E}
\def\d{\text{\rm{d}}} \def\bb{\beta} \def\aa{\alpha} \def\D{\scr D}
  \def\si{\sigma} \def\ess{\text{\rm{ess}}}
\def\beg{\begin} \def\beq{\begin{equation}}  \def\F{\scr F}
\def\Ric{\text{\rm{Ric}}} \def\Hess{\text{\rm{Hess}}}
\def\e{\text{\rm{e}}} \def\ua{\underline a} \def\OO{\Omega}  \def\oo{\omega}
 \def\tt{\tilde} \def\Ric{\text{\rm{Ric}}}
\def\cut{\text{\rm{cut}}} \def\P{\mathbb P} \def\ifn{I_n(f^{\bigotimes n})}
\def\C{\scr C}      \def\aaa{\mathbf{r}}     \def\r{r}
\def\gap{\text{\rm{gap}}} \def\prr{\pi_{{\bf m},\varrho}}  \def\r{\mathbf r}
\def\Z{\mathbb Z} \def\vrr{\varrho} \def\ll{\lambda}
\def\L{\scr L}\def\Tt{\tt} \def\TT{\tt}\def\II{\mathbb I}
\def\i{{\rm in}}\def\Sect{{\rm Sect}}  \def\H{\mathbb H}
\def\M{\scr M}\def\Q{\mathbb Q} \def\texto{\text{o}} \def\LL{\Lambda}
\def\Rank{{\rm Rank}} \def\B{\scr B} \def\i{{\rm i}} \def\HR{\hat{\R}^d}
\def\to{\rightarrow}\def\l{\ell}\def\iint{\int}
\def\EE{\scr E}\def\Cut{{\rm Cut}}
\def\A{\scr A} \def\Lip{{\rm Lip}}\def\kk{\kappa}\def\a{\mathbf a}
\def\BB{\scr B}\def\Ent{{\rm Ent}}\def\L{\scr L}\def\vol{{\rm vol}}\def\k{{\mathbf k}}
\maketitle

\begin{abstract} Two-sided estimates for higher order eigenvalues are presented for a class of non-local Schr\"odinger operators by using the jump rate and the growth of the potential.   For instance, let $L$ be the generator of a L\'evy process with L\'evy measure
$\nu(\d z):= \rr(z)\d z$ such that $\rr(z)=\rr(-z)$ and $$c_1    |z|^{-(d+\aa_1)}\le \rr(z)\le c_2|z|^{-(d+\aa_2)},\ \ |z|\le \kk $$ for some constants $\kk, c_1,c_2>0$ and $\aa_1,\aa_2\in (0,2),$
and let   $c_3|x|^{\theta_1} \le V(x)\le c_4|x|^{\theta_2}$ for some constants $\theta_1,\theta_2, c_3,c_4>0$ and large $|x|$. Then the eigenvalues $\ll_1\le \ll_2\le\cdots \ll_n\le \cdots $ of $-L+V$ satisfies the following two-side estimate: for any $p>1$, there exists a constant $C>1$ such that
$$C n^{\ff{\theta_2\aa_2}{d(\theta_2+\aa_2)}}\ge \ll_n \ge C^{-1} n^{\ff{\theta_1\aa_1}{d(\theta_1+\aa_1)}},\ \ n\ge 1.$$ When $\aa_1$ is variable,   a better lower bound estimate is derived.
\end{abstract} \noindent
 \vskip 2cm

\section{ Introduction}

Spectral asymptotics of Schr\"odinger operators and more general pseudo-differential operators is since the fundamental work of H. Weyl \cite{Weyl} a rather classical topic. The monographs of M. Reed and B. Simon \cite{Reed/Simon IV} as well as \cite{Cycon et al.} give a certain overview about techniques and results until the mid 1980s mainly based on functional analytical considerations.

Our understanding of spectral problems has undergone substantial changes with the emergence of micro-local analysis which in particular takes the symplectic structure of the co-tangent bundle and the Hamiltonian system associated with the (principal) symbol of the pseudo-differential operator under investigation into account. Of course this approach is most interesting for operators in mathematical physics, but it requires certain regularity properties of the symbols and the potentials, e.g. smoothness, the existence of a principal symbol, etc. In addition to the treatment in L. H\"ormander \cite{G}  we refer to the monographs of M.A. Shubin \cite{Shubin} and V.Ya. Ivrii \cite{Ivrii1}, as well as to the related work by C.L. Fefferman \cite{Fefferman}. In \cite{Ivrii2} also a short historical account is provided. More recently, certain pseudo-differential operators which allow certain anisotropic behaviour are treated in \cite{Nicola/Rodino}.

A further shift of the topic was given when at the relation of heat semigroups to spectral geometry was looked in more detail, in particular in the context of the Atiyah-Singer index theorem, we refer just to the monographs of P.B. Gilkey \cite{Gilkey1} and \cite{Gilkey2}. The interplay of probability theory and positivity preserving one-parameter semi-groups with spectral theory lead to a clarification of the role played by functional inequality,  we refer to the monographs  \cite{Bakry}   and \cite{W05}.

In this paper, we investigate asymptotics of eigenvalues for the Schr\"odinger type operator
$L-V $ by using  the intrinsic super Poincar\'e inequality introduced in \cite{W02}. Here,
  $L $ is the generator of a  jump  Markov process, and $V$ is a measurable function.   Comparing with Weyl's asymptotic formulae derived in the literature where regularities of $L$ and $V$ are required,  we only use   conditions on the finite-range jump rate  and the growth of the potential  $V$,  due to the stability of functional inequalities under rough perturbations.

  There is a lot of work done on non-local Schr\"odinger operators related to certain L\'evy or L\'evy-type processes, in particular the case of symmetric stable processes is well studied including the situation when it is restricted in some sense to a sub-domain, and this is followed by work on Schr\"odinger operators related to subordinate Brownian motion which includes the so-called relativistic Schr\"odinger operator. For a general discussion of subordinate Brownian motion we refer to \cite{U,C}.

With respect to a study of such Schr\"odinger operators we refer to earlier papers of K. Bogdan and T. Byczkowski \cite{A,B}, the important paper by Z.-Q. Chen and R. Song \cite{E}, to the string of papers by K.Kaleta and co-authors \cite{K,L,M,N}, the work of T. Kulczycki \cite{P}, and that of M. Kwasnicki \cite{Q} as well as the papers by J. L\"orinczi and co-authors\cite{R,S}, just to mention a few contributions.

The background of our starting point differs a bit from other investigations. It is meanwhile apparent that Dirichlet form techniques and related stochastic analysis or methods from the theory of functional inequalities have lead to enormous progress in our understanding of jump-type processes, or more precisely L\'evy-type processes which should be looked at as processes having as generator a pseudo-differential operator with a negative definite symbol, we refer to the recent survey initiated by R. Schilling \cite{C}. However, it seems that certain problems are out of the reach of our current techniques, for example we lack a geometric interpretation of transition densities as we do have for local, sub-elliptic operators, or when discussing Feynman-Kac formulae and a possible semi-classical asymptotic we essentially do not have a $``$classical'' counterpart.

In \cite{H}, see also \cite{D}, a suggestion was made to approach the first problem. In more recent work it was started to develop the Hamiltonian dynamics behind certain L\'evy processes, i.e. to consider the symbol of a generators as Hamiltonian function. A good starting point is to look at $H(q,p) = \psi(p) + V(q)$ where $\psi $ is a certain convex, coercive negative definite function of class $C^1$ and V is a suitable potential, see \cite{J,T}. As substitute for the harmonic oscillator it was proposed to consider $H(q,p) = \psi(p) + \psi^*(q)$ where $\psi^*$ is the conjugate convex function (Legendre transform) of $\psi$. In this context now arises the question whether the study of the symbol $a(x,\xi) = \psi(\xi) + \psi^*(x)$ on the co-tangent bundle will allow us to derive for example spectral results as they are known for elliptic differential operators, we only refer to the monograph \cite{G} of L. H\"ormander.

Since functional inequalities also can lead to information on eigenvalues, see for instance \cite{W05}, it was natural to raise the question posed in the beginning. Once the problem was laid out, it was possible to employ techniques and results from the theory of functional inequalities as developed in \cite{W00,W02,WW}   and to come up with some eigenvalue asymptotic, see Theorem 1.3 below. Of special interest was to include an example constructed with the help of a non-smooth, i.e. not even $C^2$, convex and coercive negative definite function which is anisotropic, i.e. not a subordinate Brownian motion, see    Examples 2.1 and  2.2 below.

\

In the next section we introduce  two results for $\aa$-stable like Schr\"odinger operators,  where the first allows the power $\aa$ varies and the second compares $\aa$ with a constant. Concrete examples are addressed to illustrate these results, which are proved in Sections 3 and 4 respectively. 

\section{Main results and example}

 By using the intrinsic super Poincar\'e inequality introduced in \cite{W02}, the compactness of Schr\"odinger semigroups have been investigated in \cite{WW} under an abstract framework.
 Let $E$ be a Polish space with a $\si$-finite measure $\mu$. Let $(L_0,\D(L_0))$ be a symmetric Dirichlet operator on $L^2(\mu)$ such that the associated Markov semigroup $P_t^0$  is ultracontractive, i.e. $\|P_t^0\|_{L^1(\mu)\to L^\infty(\mu)}<\infty$ for $t>0$. We consider the Schr\"odinger operator $L_V:=L-V$, where $V$ is a locally integrable nonnegative measurable function on $E$
 such that $\D :=\{f\in \D(L_0):\ \mu(Vf^2)<\infty\}$ is dense in $L^2(\mu)$. Then the   Friedrichs extension   $(L_V,\D(L_V))$ of $(L_V,\D)$  is a negatively definite self-adjoint operator generating a sub-Markov semigroup $P_t^V$ on $L^2(\mu)$.  The following result follows from \cite[Theorem 1.1]{WW} which indeed applies   to a more general setting, see also \cite{S} for an alternative proof when $L_0=\DD$ on $\R^d$.

 \beg{thm}[\cite{WW}]  \label{T1.0} If $\mu( V\le r)<\infty$ for $r\ge 0$, then for any $t>0$, $P_t^V$ is compact  in $L^2(\mu)$.  \end{thm}

It is well know that the compactness of  $P_t^V$ is equivalent to the absence of the essential spectrum of $L_V$. In this case $-L_V$ has purely discrete spectrum and all eigenvalues, including multiplicities,  can be listed as
$$0\le \ll_1\le\ll_2\le\cdots\le\ll_n\le\cdots$$ with $\lim_{n\to\infty}  \ll_n=\infty.$
In this paper we aim to investigate upper and lower bound estimates of  $\ll_n$  for $L_0$ being a   non-local  symmetric operator on $L^2(\R^d)$.

\

Let $J: \R^d\times\R^d\to [0,\infty]$ be measurable such that
\beg{enumerate} \item[$(A)$] $J(x,y)=J(y,x)$ and
\beq\label{C0} \sup_{x\in \R^d} \int_{\{|z|\le 1\}} |z|\cdot|J(x, x+z)-J(x,x-z)|\d z<\infty,\end{equation}
 \beq\label{C1}  \sup_{x\in\R^d} \int_{\R^d} (|z|^2\land 1) J(x,x+z)\d z<\infty.\end{equation}
 \end{enumerate}
Let $V\ge 0$ be a  locally integrable function on $\R^d$. Then the following non-local Schr\"odinger operator is well defined in $L^2(\R^d)$  for $f\in C_0^\infty(\R^d)$:
\beg{align*} L_{J,V}f(x):= &\int_{\R^d} \big\{f(x+z)-f(x) -\<\nn f(x), z\>1_{\{|z|\le 1\}}\big\}J(x,x+z)\d z\\
&+ \ff 1 2 \int_{\{|z|\le 1\}} \<\nn f(x), z\> \big(J(x,x+z)-J(x,x-z)\big)\d z -(Vf)(x),\ \ x\in\R^d.\end{align*}
Moreover,  we have the following integration by parts formula
\beg{align*} &\int_{\R^d} (fL_{J,V}g)(x)\d x =-\scr E_{J,V}(f,g),\ \ f,g\in C_0^\infty(\R^d),\\
&\scr E_{J,V}(f,g):= \int_{\R^d\times\R^d} \big(f(x)-f(y)\big)\big(g(x)-g(y)\big)J(x,y)\d x\d y +\int_{\R^d} (Vfg)(x)\d x.\end{align*} Therefore, the form $(\scr E_{J,V}, C_0^\infty(\R^d))$ in $L^2(\R^d)$ is closable  and the closure $(\scr E_{J,V}, \D(\scr E_{J,V}))$ is a symmetric Dirichlet form, the associated generator  $(L_{J,V}, \D(L_{J,V}))$  is the Friedrichs extension  of  $(L_{J,V}, C_0^\infty(\R^d))$. Let $P_t^{J,V}$ be the associated sub-Markov semigroup.

According to \cite[Theorem 1.1]{BC}, under a suitable lower bound condition on $J$, the Markov semigroup $P_t^{J,0}$ generated by $L_{J,0}$ is ultracontractive with respect to  the Lebesgue measure. By Theorem \ref{T1.1},
if $V(x)\to \infty$ as $|x|\to\infty$, the essential spectrum of $L_{J,V}$ is empty. Let $\ll_1\le\ll_2\le \cdots $ be all eigenvalues of $-L_{J,V}$. We will estimate $\ll_n$ in terms of $\aa_1,\aa_2$ in
$(A2)$ and the growth of $V(x)$ as $|x|\to\infty$.   Obviously, $\EE_{J,V}(f,f)=0$ if and only if $f=0$. This implies
  $\ll_1>0$.

To estimate $\ll_n$ from below,  we will  use  the following  intrinsic super Poincar\'e inequality introduced in \cite{W02}:
\beq\label{SI} \int_{\R^d} f^2(x)\d x\le s \scr E_{J,V}(f,f)+ \bb(s) \bigg(\int_{\R^d} |f\phi |(x)\d x\bigg)^2,\ \ s>0, f\in C_0^\infty(\R^d),\end{equation}
where $\bb: (0,\infty)\to (0,\infty)$ is a decreasing function, and $\phi\in L^2(\R^d)$ is a reference function. Let
$$\bb^{-1}(r)=\inf\{s\ge 0: \bb(s)\le r\},\ \ r>0,$$ where $\inf\emptyset =\infty$ by convention.
The following  result is essentially due to   \cite{W02}, see Section 2 for a complete proof.

\beg{thm}[\cite{W02}] \label{T1.1} Let $\phi\in L^2(\R^d)$ be positive such that $P_t^{J,V}\phi\le \e^{\ll t}\phi$ holds for some $\ll\in\R$ and all $t\ge 0$.   If $\eqref{SI}$ holds for some $\bb$ with $\bb(\infty):=\lim_{s\to\infty} \bb(s)=0$ and
$$\LL(t):= \int_t^\infty \ff{\bb^{-1}(r)}{r}\d r<\infty,\ \ t>0,$$
then  for any $\vv\in (0,1)$, there exists a constant $c>0$ such that
\beq\label{LL*}\ll_n\ge  \ff {c }{\LL( \vv n)}, \ \ \ n\ge 1.\end{equation}\end{thm}

According to this result, to derive sharp lower bound of $\ll_n$, we need  to prove the inequality  \eqref{SI} for  as small as possible $\bb$.   Intuitively,  to establish   \eqref{SI} with smaller $\bb$, we should take   larger $\phi\in L^2(\R^d)$. In this spirit, reasonable  choices of $\phi$  will be
$\phi(x)= (1+|x|^2)^{-p}$ for some constant $p> \ff d 4$, or $\phi(x)=\varphi_k(|x|)$ for some $k\in\mathbb Z_+$ and $p>1$, where
\beq\label{PK}\varphi_k(s):= (1+s^2)^{-\ff d 4}\cdot\Big(\{ \log^{k+1}(\e^{k+1}+s^2)\}^{\ff p 2} \cdot \prod_{i=1}^k\ss{\log^{i} (\e^i+s^2)}\Big)^{-1},\ \ s\ge 0,\end{equation} $\log^{k+1}:=\log^k\log $ for $k\ge 1$, and $\prod_{i=1}^k:=1$ if $k=0$.     In general, we consider $\phi(x)=\varphi(|x|)$ for   $\varphi\in\scr S,$ the class of decreasing functions $\varphi\in C^2( [0,\infty);\R_+)$ such that
\beg{enumerate}\item[(i)] $\int_0^\infty s^{d-1}\varphi(s)^2\d s <\infty; $\item[(ii)]\ There exists a constant $c>0$ such that
$$\sup_{r\le 1+s}\Big( |\varphi'(r)| (r+r^{-1})+ |\varphi''(r)|\Big)+\varphi(s/2) \le c \varphi(s),\ \ s\ge 0.$$\end{enumerate}
By establishing the intrinsic Poincar\'e inequality \eqref{SI} for $\phi(x):= \varphi(|x|)$, we will derive the following main result of the paper.

\beg{thm}\label{T1.2} Let $J$ satisfy $(A)$.
  Assume that for some constant $c>0$ and symmetric function $\aa \in C(\R^d\times\R^d;(0,2))$ there holds
\beq\label{LB} J(x,y)\ge \ff{c_1}{|z|^{d+\aa(x,y)}} 1_{\{|x-y|\le \kk\}},\ \ x,y\in\R^d.\end{equation}
Let $\a(r)= \inf_{|x|\lor |y|\le r} \aa(x,y)$ and
$\Phi(R):= \inf_{|x|\ge R} V(x)\uparrow \infty$ as $R\uparrow\infty$.  For $\varphi\in\scr S$ and  $\Phi^{-1}(r)=\inf\{s\ge 0: \Phi(s)\ge r\},\ r\ge 0,$ let
$$\GG(r)= \inf\big\{s>0: s^{d/\a(\Phi^{-1}(2s^{-1})+1)}\varphi\big(\kk+\Phi^{-1}(2s^{-1})\big)^{2} \ge r^{-1}\big\},\ \ r>0.$$
If
$$\ll(t):= \int_t^\infty \ff{\GG(r)}{r}\d r<\infty,\ \ t>0,$$
then there exist  constants $\dd_1,\dd_2>0$ such that
\beq\label{LL0}\ll_n\ge   \ff{\dd_1}{\ll(\dd_2 n)},\ \ \ n\ge 1.\end{equation}
\end{thm}

\paragraph{Remark 2.1.} Condition \eqref{C0} in $(A)$   will be  only used to verify the condition
\beq\label{P0} P_t \phi(x)\le \e^{ct} \phi(x),\ \ x\in\R^d, t\ge 0,\end{equation} where $P_t$ is the Markov for the jump process with jump kernel $J(x,y)$,  $\phi\in \scr S$, and $c>0$ is a constant depending on $\phi$. If the heat kernel of $P_t$  satisfies the upper bound estimate
\beq\label{P1}p_t(x,y)\le c t(t^{\ff 1 \aa}+ |x-y|)^{-(d+\aa)},\ \ x,y\in \R^d, t>0\end{equation}  for some constants $c>0$ and $\aa\in (0,2)$,
then $P_t\phi\le CP_t^\aa \phi$ holds for some constant $C>0$, where $P_t^\aa$ is the semigroup of the jump process with jump kernel $J_\aa(x,y):=|x-y|^{-(d+\aa)}$ which trivially satisfies condition \eqref{C0}. According to the proof of  Lemma \ref{L2} below for $J_\aa$ replacing $J$, we have
$$P_t^\aa \phi\le \e^{\ll t} \phi$$ for some constant $\ll>0$, so that \eqref{P0} follows. Therefore, under the heat kernel estimate \eqref{P1}, we can drop \eqref{C0} from assumption $(A)$ in Theorem \ref{T1.2}.

\

Consider the following stable-like jump kernel $J$ satisfying
\beq\label{*P0}J(x,y)1_{\{|x-y|\le\kk\}}= \ff{n(x,y)}{|x-y|^{d+\aa(x,y)}} 1_{\{|x-y|\le \kk\}},\ \ x,y\in \R^d,\end{equation} where $\kk>0$ is a constant and
$$n, \aa: \R^d\times \R^d\to [0,\infty)$$ are measurable and symmetric, such that
\beg{enumerate}
\item[(a)] There exists a constant $\vv\in (0,1)$ such that $\vv\le n(x,y)\le \vv^{-1}, x,y\in\R^d;$
\item[(b)] There exists constants $\aa_2\in (0,2)$ such that $0< \aa(x,y)\le \aa_2, x,y\in\R^d;$
\item[(c)] $\sup_{x\in\R^d} \int_{\{|z|\le 1\}} \ff{|n(x,x+z)-n(x,x-z)|}{|z|^{d+\aa_2-1}}\d z<\infty.$\end{enumerate}
Then assumptions $(A)$ holds, so that Theorem \ref{T1.2} applies.

For instance, we have the following concrete example.

\paragraph{Example 2.1.} Let $J$ be in $\eqref{*P0}$ with
 $$\aa(x,y) =\aa_0+\ff {\bb_1} {[\log (\bb_2 + |x|+|y|)]^{\ff 1 2}},$$ where $\aa_0\in (0,2)$ and $\bb_1>0,\bb_2>1$ such that $\ff{\bb_1}{(\log \bb_2)^{\ff 1 2}}\in (0, 2-\aa_0).$
 If $    V(x)  \ge c  |x|^\theta  $ holds  for some constants $c,\theta>0$,  then for any $\dd\in \big(0, \ff{d\bb_1\ss\theta}{\aa_0^2}\big(\ff{d(\aa_0+\theta)}{\aa_0\theta}\big)^{\ff 3 2}\big)$ there exists a constant $c(\dd)>0$ such that
\beq\label{PO}   \ll_n \ge  c(\dd) n^{\ff{\theta\aa_0}{d(\theta+\aa_0)}} \exp\Big[\dd \ss{\log n}\Big],\ \ n\ge 1.\end{equation}

\beg{proof} We may take $\Phi(r)= cr^\theta$ for large $r>0$ and $\a(r)= \aa_0+ \ff{\bb_1}{\{\log(\bb_2+2r)\}^{\ff 1 2}}$ for $r>0$. Taking $\varphi (r)= (1+ r^2)^d$, for   $s\in (0,1)$ we have
\beg{align*} &\Phi^{-1} (2s^{-1}) \le c_1 s^{-\ff 1 \theta},
&\a( \Phi^{-1} (2s^{-1}) +1) \ge \aa_0+ \ff{\bb_1\ss\theta}{\ss{ \log s^{-1}}} + {\rm o}\big((\log s^{-1})^{-\ff 1 2}\big).\end{align*}
So, there exist  constants $c_1,c_2>0$ such that
\beg{align*} s^{d/\a(\Phi(2s^{-1})+1)}&\le  s^{\ff d \aa_0  - \ff{d\bb_1\ss\theta}{\aa_0^2\ss{\log s^{-1}}+c_1}}\\
&\le  c_2 s^{\ff d \aa_0}
\exp\bigg[-\ff{d\bb_1\ss\theta}{\aa_0^2}\ss{\log s^{-1}} \bigg],\ \ s\in (0,s_0\land \e^{-1}].\end{align*}
Let   $\varphi(r)= (1+r^2)^{-\ff d 4} \{\log(\e +r^2)\}^{-2}$. Then $\varphi\in \scr S$ and  for any constant $\dd'\in (0, d\bb_1\ss\theta\aa_0^{-2})$, there exists constant $c_1(\dd')>0$ such that
$$s^{d/\a(\Phi(2s^{-1})+1)} \varphi\big(\kk+\Phi^{-1}(2s^{-1})\big)^{2}
 \le c(\dd')s^{\ff{d(\theta+\aa_0)}{\theta\aa_0} }\exp\Big[-\dd'  (\log s^{-1})^{\ff 1 2}\Big],\ \ s\in (0,s_0\land \e^{-1}].$$
Then there exists a constant $c_2(\dd')>0$ such that
$$\GG(r) \le c_2(\dd') r^{-\ff{\theta\aa_0}{d(\theta+\aa_0)} } \exp\bigg[- \dd'\Big(\ff{d(\theta+\aa_0)}{\theta\aa_0}\Big)^{\ff 3 2}
\ss{\log r}\bigg],\ \ r\ge 1,$$
so that for some constant $c_3(\dd')>0$,
$$\ll(t):=\int_t^\infty\ff{\GG(r)}r \d r \le c_3(\dd')  t^{-\ff{\theta\aa_0}{d(\theta+\aa_0)} } \exp\bigg[- \dd'\Big(\ff{d(\theta+\aa_0)}{\theta\aa_0}\Big)^{\ff 3 2}\ss{\log t} \bigg],\ \ t\ge 1.$$
Since $\dd'\in (0, d\bb_1\ss\theta\aa_0^{-2})$ is arbitrary, this  implies   \eqref{PO} for $\dd\in \big(0, \ff{d\bb_1\ss\theta}{\aa_0^2}\big(\ff{d(\aa_0+\theta)}{\aa_0\theta}\big)^{\ff 3 2}\big)$.
\end{proof}
\

When  $J$ is comparable  with the  $\aa$-stable kernel of finite range, we have the following sharp result. But in general the lower bound estimate \eqref{LL2} may be less sharp than that given in Theorem \ref{T1.2}. For instance, applying Theorem \ref{T1.3}(1) to Example 2.2 one only derives
 $$\ll_n\ge c n^{\ff{\theta\aa_0}{d(\theta+\aa_0)}},\ \ n\ge 1$$ which is less sharp than \eqref{PO}.

\beg{thm}\label{T1.3} Let $J$ satisfy $(A)$ and let $\theta, \kk>0,\aa\in (0,2)$ be constants. \beg{enumerate} \item[$(1)$]  If there exists a constant $c>0$ such that    $V(x)\ge c|x|^\theta $ for   large $|x|$, and
\beq\label{LB} J(x,y)\ge \ff{c}{|z|^{d+\aa}} 1_{\{|z|\le \kk\}},\ \ x,y\in\R^d,\end{equation}   then    there exists a constant $\dd>0$ such that
\beq\label{LL2}\ll_n\ge \dd n^{\ff{\theta\aa}{d(\theta+\aa)}},\ \ n\ge 1. \end{equation}
\item[$(2)$] If there exists a constant $c>0$ such that $V(x)\le c |x|^\theta  $ for   large $|x|$, and
\beq\label{UP} J(x,y)1_{\{|x-y|\le \kk\}}\le \ff{c }{|x-y|^{d+\aa }},\ \ x,y\in\R^d,\end{equation}
   then there exists a constant $\dd>0$ such that
\beq\label{LL2'} \ll_n\le \dd n^{\ff{\theta \aa }{d(\theta +\aa )}},\ \ n\ge 1.\end{equation}\end{enumerate}
\end{thm}

 We now apply Theorem \ref{T1.3}  to specific models induced by symbols of pseudo differential operators. Let
 $$\psi(\xi)=\int_{\R^d} (1-\cos x\cdot \xi) \nu(\d x) ,\ \ \ \xi\in\R^d$$ for a L\'evy measure $\nu$ with $\nu(1\land|\cdot|^2)<\infty$. Typical examples of $\psi$ include
 $\psi(\xi)=|\xi|^\aa$ for $\aa\in (0,2)$, or in general
 $$\psi(\xi)= \sum_{i=1}^m c_i \Big(\sum_{j=1}^d |\xi_j|^{\aa_{ij}}\Big)^{\bb_j}$$
 for some constants $m\in \mathbb N$ and $ c_i, \aa_{ij}, \bb_j>0$ with $\bb_j\max_{i} \aa_{ij}<2$.
 Let $\psi(D)$ be the pseudo differential operator induced by $\psi$, see for instance \cite{Jacob}. We consider   the  Schr\"odinger operator
$$L_{\psi,V}:= -\psi(D)+V$$ for a nonnegative potential $V$.

\paragraph{Example 2.2.}  Assume that
\beq\label{PS} c_1 |\xi|^{\aa} \le \psi(\xi)\le c_2 (|\xi|^{\aa}+ |\xi|^{\aa'}),\ \ \xi\in \R^d\end{equation} holds for some constants  $2>\aa\ge \aa'>0$ and $c_1,c_2>0$. If
\beq\label{VV}  c_3|x|^\theta\le V(x)\le c_4  |x|^\theta \ \ \text{for\ large}\ |x|,\end{equation} holds for some constants $\theta, c_3,c_4>0$, then \beq\label{EG}\dd_1 \dd n^{\ff{\theta \aa }{d(\theta +\aa )}}\le \ll_n\le \dd_2 n^{\ff{\theta \aa }{d(\theta +\aa )}},\ \ n\ge 1\end{equation} holds for some constants $\dd_1,\dd_2>0.$

\beg{proof}  It is well known that $L_{\psi,V}= L_{J,V}$ for some jump rate $J$ with $J(x,y)=J(0,x-y)=J(0,y-x)$ so that $(A1)$ holds, and  by \eqref{PS},   there exists a constant $C>1$ such that
$$ \ff{ 1_{\{|x-y|\le 1\}}}{C|x-y|^{d+\aa}}  \le J(x,y)1_{\{|x-y|\le 1\}}\le  \ff{ C 1_{\{|x-y|\le 1\}}}{|x-y|^{d+\aa}},\ \ x,y\in \R^d.$$
   Then the proof is finished by
Theorem \ref{T1.3}.
\end{proof}

\paragraph{Example 2.3.} Consider, for instance, $\psi(\xi)= (|\xi_1|^{r_1} + |\xi_2|^{\bb_1} )^{\gg_1} + (|\xi_1|^{r_2} + |\xi_2|^{\bb_2} )^{\gg_2}$ on $\R^2$ for some constants
$r_i,\bb_i,\gg_i>0$ such that $r_1\gg_1=\bb_2\gg_2, r_2\gg_2=\bb_1\gg_1\in (0,2).$  Then condition \eqref{PS} holds for
$$\aa_1:=\max \{r_1\gg_1, \bb_1\gg_1\}, \ \ \aa_2 :=\min \{r_1\gg_1, \bb_1\gg_1\}.$$  By Theorem \ref{T1.3}, if   \eqref{VV}  holds  then  eigenvalues $\{\ll_n\}_{n\ge 1} $ of $-L_{\psi,V}$  satisfies
 \eqref{EG}  for some constants $\dd_1,\dd_2>0.$

Moreover, for $r_1\gg_1>r_2\gg_2>1$ the function $\psi$ is coercive, convex, negative definite and satisfies
$$K_0 |\xi|^{r_1\gg_1}\le\psi(\xi)\le K_1 \big(|\xi|^{r_1\gg_1} +|\xi|^{r_2\gg_2}\big)$$ for some constants $K_0,K_1>0.$ The convex conjugate function $\psi^*$ is again coercive, i.e. $\lim_{|\xi|\to\infty} \ff{\psi^*(\xi)}{|\xi|} = \infty$, and satisfies $\psi^*(\xi)\le K_2|\xi|^{(r_1\gg_1)^*}$ for $(r_1\gg_1)^*:= \ff{r_1\gg_1}{r_1\gg_1-1}.$ With the help of \cite[Theorem 3 on page 87]{F}, we deduce that $\psi^*\ge 0$. Hence we may apply Theorem \ref{T1.3} to $\psi(D)+\psi^*(x).$

\section{Proof of Theorem \ref{T1.2}  }

We first prove Theorem \ref{T1.1} using results in \cite{W02}.

\beg{proof}[Proof of Theorem \ref{T1.1}] Without loss of generality, we may and do assume that $\mu_\phi(\d x):= \phi(x)^2\d x$ is a probability measure.  Simply denote  $P_t=P_t^{J,V}=\e^{tL_{J,V}},$   the (sub)-Markov semigroup generated by $L_{J,V}.$   We consider the following symmetric semigroup
$P_t^{\phi}$ on $L^2(\mu_{\phi})$:
$$P_t^{\phi}f(x):= \phi^{-1} P_t(f\phi),\ \ \ f\in L^2(\mu_{\phi}),\ \ t\ge 0.$$
According to \cite[Theorem 3.3(1)]{W02}  for $\inf\bb=0$, $P_t^{\phi}$ has a symmetric heat kernel $p_t^{\phi}(x,y)$ with respect to $\mu_{\phi}$, i.e.
$$P_t^{\phi} f(x)= \int_{\R^d} p_t^{\phi}(x,y) f(y) \phi(y)^2 \d y,\ \ \ f\in L^2(\mu_{\phi}), t>0,$$
such that
$$\sup p_t^{\phi}= \|P_t^{\phi}\|_{L^1(\mu_{\phi})\to L^\infty(\mu_{\phi})} \le \e^{\ll  t} \LL^{-1}(\vv t)^2,\ \ t>0, \vv\in (0,1)$$
where $\LL^{-1}(t):=\inf\{r>0: \LL(r)\le t\}<\infty,\ \ t>0.$ Since  $\LL^{-1}$ is continuous, by letting $\vv\to 0$ we obtain
$p_t^{\phi}\le \e^{\ll t}   \LL^{-1}(t)^2$.
Consequently, the heat kernel $p_t$ of $P_t$ with respect to the Lebesgue measure has the upper bound estimate
$$p_t(x,y)\le \phi(x)\phi(y) \e^{\ll t} \LL^{-1}(t),\ \ t>0,\ x,y\in\R^d.$$
In particular, $\int_{\R^d} p_t(x,x) \d x\le  \e^{\ll t} \LL^{-1}(t),\ t>0.$
Combining this with \cite[Theorem 2.4]{W02}, we obtain
$$\ll_n\ge \sup_{t>0} \ff 1 t \log\ff {n\e^{-\ll t}}{\LL^{-1}(t)},\ \ n\ge 1.$$  With $t=\LL(\vv n)$ this implies
\beq\label{GD}\ll_n\ge \ff {\log (\vv^{-1} \e^{-\ll \LL(\vv n)})}{\LL(\vv n)},\ \ n\ge 1.\end{equation}
Since $\LL(r)\to 0$ as $r\to\infty$, there exists  $n_0\ge 1$  such that $c_0:=\vv^{-1}\e^{-\ll \LL(\vv n_0)}>1$. By the decreasing monotonicity of   $\LL,$ we have
$\vv^{-1} \e^{-\ll(\vv n)} \ge c_0>1$ for $n\ge n_0$. So,   \eqref{GD} implies  \eqref{LL*} for
$c:= \log c_0>0$ and $n\ge n_0$. Combining this with $\ll_1>0$, we conclude that  \eqref{LL*}  holds for   $c:=\min\{\log c_0, \ll_1\LL(\vv n_0)\}>0$ and all $n\ge 1.$

 \end{proof}

To prove Theorem \ref{T1.2} using Theorem \ref{T1.1}, we first verify $P_t^{J,V}\phi\le \e^{\ll t}\phi$ for $\phi:=\varphi(|\cdot|)$ and   $\ll>0$.

\beg{lem}\label{L2} For any positive $\varphi\in C^2([0,\infty)$ satisfying condition {\rm (ii)},  there exists a constant $\ll\ge 0$ such that
$$P_t^{J,V}\varphi(|\cdot|)(x)\le \e^{\ll t} \varphi(|\cdot|)(x),\ \ \ t\ge 0, x\in \R^d.$$\end{lem}
\beg{proof} Let $\phi(x)= \varphi(|x|)$. Then condition (ii) implies
\beq\label{Y1} \beg{split}&\sup_{|z|\le 1} |z|^{-2} \big|\phi(x+z)-\phi(x)-\<\nn\phi(x), z\>  \big|\le \sup_{|z|\le 1} \|\Hess_\phi(x+z)\|\\
&\le \sup_{r\in (0, 1+|x|]} \Big(\ff{|\varphi'(r)|}{r}+|\varphi''(r)|\Big)\le c_1 \phi(x),\ \ \ x\in\R^d\end{split}\end{equation}
for some constant $c_1>0$. Next, if $|x+z|\ge \ff 1 2|x|$, then       (ii) implies
$$\phi(x+z)-\phi(x)\le \phi(|x|/2)\le c\phi(x)$$ for some constant $c>0$; while if $|x+z|<\ff 1 2|x|$ then $\ff 1 2 |x|\le |z|\le \ff 3 2|x|$, so that
(ii) gives
$$\phi(x+z)-\phi(x)\le |z|\int_0^1|\varphi'(|x+sz|)|\d s \le c\sup_{r\in (0, \ff 5 2 |x|]} r |\varphi'(r)|\le c' \phi(x)$$
for some constants $c,c'>0$. Combining these with \eqref{Y1} we conclude that
$$\big|\phi(x+z)-\phi(x)-\<\nn\phi(x), z\>1_{\{|z|\le 1\}}\big|\le c_2 (|z|^2\land 1)\phi(x)$$ holds for some constant $c_2>0$. Therefore,
it follows from condition \eqref{C1} in $(A)$  that
\beq\label{Y2}
  \int_{\R^d} \big(\phi(x+z)-\phi(x)-\<\nn\phi(x),z\>1_{\{|z|\le 1\}}\big)J(x,x+z)\d z\le c_3 \phi(x),\ \ \ x\in\R^d \end{equation} holds for some constant $c_3>0$.
Moreover, by (i) and condition \eqref{C0} in $(A)$, we have
\beg{align*}&\bigg|\int_{\{|z|\le 1\}} \<\nn\phi(x),z\>\big(J(x,x+z)-J(x,x-z)\big)\d z\bigg|\\
&\le  c_4\phi(x) \int_{\{|z|\le 1\}} |z|\cdot \big|J(x,x+z)-J(x,x-z)\big|\d z\le c_5\phi(x),\ \ x\in\R^d\end{align*}
for some constants $c_4,c_5>0$. This together with \eqref{Y2} yields
\beg{align*}L^{J,0}\phi(x) :=  &\int_{\R^d} \big(\phi(x+z)-\phi(x)-\<\nn\phi(x),z\>1_{\{|z|\le 1\}}\big)J(x,x+z)\d z\\
&+ \int_{\{|z|\le 1\}} \<\nn\phi(x),z\>\big(J(x.x+z)-J(x,x-z)\big)\d z\le \ll \phi(x)\end{align*}
for some constant $\ll\ge 0$ and all $x\in\R^d$. Therefore, letting  $X_t^x$ be the jump process  with jump rate $J$ starting at $x$, we obtain
$$ P_t^{J,0}\phi(x):= \E[\phi(X_t^x)]\le \e^{\ll t} \phi(x),\ \ t\ge 0, x\in \R^d.$$ By Feynman-Kac formula   and $V\ge 0$, we conclude that
$$P_t^{J,V} \phi(x)= \E\big[\phi(X_t^x) \e^{-\int_0^t V(X_s^x)\d s}\big]\le P_t^{J,0}\phi(x)\le \e^{\ll t} \phi(x),\ \ t\ge 0, x\in \R^d.$$
\end{proof}

\beg{proof}[Proof of Theorem \ref{T1.2}]    According to Theorem \ref{T1.1} and Lemma \ref{L2},   it suffices to estimate the rate function $\bb$ in \eqref{SI}.

Let $\phi(x)=\varphi(|x|)$.
We first prove   the following local super Poincar\'e inequality
\beq\label{*Q1}  \int_{B(0,r)} f^2(x)
d x \le s \scr E_{J,V}(f,f) +c\aa(r,s) \bigg(\int_{B(0,r+1)} (|f|\phi)(x)\d x\bigg)^2,\ \ s>0, r\ge 1,\end{equation}  for some constant $c>0$, where
\beq\label{Q1} \aa(r,s):=  \ff{c(1+s^{-\ff d{\a(r+1)}})}{  \varphi(r+1)^2}. \end{equation}
Indeed, for any $f\in L^2(\R^d)$ and $t\in (0,\kk\land 1]$, let
$$f_t(x)= \ff 1 {|B(0,t)|} \int_{B(x,t)} f(y)\d y,\ \ x\in\R^d.$$
According to step (ii) in the proof of \cite[Proposition 3.5]{CW16}, we have
\beq\label{*Q2}\beg{split}  &\int_{B(0,r)} f^2(x)\d x \le 2 \int_{B(0,r)} |f(x)-f_t(x)|^2 \d x + 2 \int_{B(0,r)} f_t^2(x)\d x \\
&\le \ff 2 {|B(0,t)|} \bigg\{ \int_{B(0,r)\times B(x,t)} |f(x)-f(y)|^2 \d x\d y+ \bigg(\int_{B(0,r+t)}|f(y)|\d y\bigg)^2\bigg\}.\end{split}\end{equation}
Since $t\le\kk\land 1$, by \eqref{LB} for $(x,y)\in B(0,r)\times B(x,t)$ we have
$$J(x,y) \ge \ff{c_1} {|x-y|^{d+\aa(x,y)}} \ge \ff{c_2}{t^{d+\a(r+1)}},$$ so that
\beq\label{*Q3} \int_{B(0,r)} |f(x)-f_t(x)|^2 \d x \le c_2^{-1} t^{d+\a(r+1)} \EE_{J,V}(f,f).\end{equation}
Moreover, since $\phi(x)=\varphi(|x|)$ is decreasing in $|x|$, it holds that
$$\int_{B(0,r+t)}|f(y)|\d y\le \ff 1 {\varphi(r+1)} \int_{B(0,r+t)}|f\phi(y)|\d y,\ \ t\in (0, 1\land\kk].$$ Combining this with  \eqref{*Q2} and \eqref{*Q3}, we may find a constant $c_3>0$ such that
 $$ \int_{B(0,r)} f^2(x)\d x \le c_3 t^{\a(r+1)}\EE_{J,V}(f,f)  + \ff{c_3} {t^d\varphi(r+1)^2} \bigg(\int_{\R^d} |f\phi|(x)\d x\bigg)^2,\ \ t\in (0,1\land \kk], r\ge 1.$$
Taking $s= c_3t^{\a(r+1)}$ and noting that $\kk^{\a(r+1)} \ge 1\land \kk^2$, we may find out  constants $c_4, s_0>0$ such that
 $$ \int_{B(0,r)} f^2(x)\d x \le s\EE_{J,V}(f,f)  + \ff{c_4s^{\ff{d}{\a(r+1)}}} {\varphi(r+1)^2} \bigg(\int_{\R^d} |f\phi|(x)\d x\bigg)^2,\ \ s\in (0,s_0], r\ge 1.$$ This implies \eqref{*Q1}.

We now    estimate $\bb$ in \eqref{SI} using \eqref{*Q1} and the growth function $\Phi$ of $V$. We first assume $\inf V>0$ then extend to the general case.

By \cite[Proposition 3.5]{CW16},   condition \eqref{LB} implies the following local super Poincar\'e inequality
 holds for

(a) Let  $K:=\ff 1 2 \inf V>0$. Then
$$\Phi(R):= \inf_{|x|\ge R, V(x)>K}V(x) =\inf_{|x|\ge R}V(x),\ \ \Theta(R):= \vol(\{x: |x|\ge R, V(x)\le K\})=0.\ \  \ R>0.$$
Combining this with   \eqref{Q1}, we may apply  \cite[Theorem 2.1]{CW16} for $r_0=0$ to obtain \eqref{SI} with
\beg{align*} \bb(s) &=\dd_3 \aa\big(\Phi^{-1}(2s^{-1}), s/2\big)\\
&\le \dd_4 \big(1+s^{-d/\a(\Phi^{-1}(2s^{-1})+1)}\big) \varphi(1+\Phi^{-1}(2s^{-1}))^{-2},\ \ s>0\end{align*}
for some constants $\dd_3, \dd_4>0$.  Since $\ll_1>0$, when $s\ge \ll_1$, \eqref{SI} holds for $\bb(s)=0$. Therefore, there exists a constant $\dd_5>0$ such that \eqref{SI} holds for
$$\bb(s):=\dd_5  s^{-d/\a(\Phi^{-1}(2s^{-1})+1)} \varphi(1+\Phi^{-1}(2s^{-1}))^{-2}=:\dd_5\tt\bb(s),\ \ s>0.$$
Then, for $\GG$ in Theorem \ref{T1.2}, we have
$$\bb^{-1}(r):=\inf\{s>0: \bb(s)\le r\}= \inf\{s>0: \tt\bb(s)\le \dd_5^{-1}r\}=  \GG(\dd_5r),\ \ r>0.$$
Thus,  $$\LL(t):= \int_t^\infty\ff{\bb^{-1}(r)}r\d r= \int_t^\infty\ff{\GG(\dd_5r)}r\d r=   \ll(\dd_5t),\ \ t>0.$$
According to Theorem \ref{T1.1} and Lemma \ref{L2},   this implies  \eqref{LL0} for some constants $\dd_1,\dd_2>0$.

(b) In general, let $\bar V= V+\ll_1$. Since $\scr E_{J,V}(f,f)\le \scr E_{J,\bar V}(f,f)$,   \eqref{SI} also holds for $\scr E_{J,\bar V}$ replacing $\scr E_{J,V}$. Since $\inf \bar V\ge \ll_1>0$,    by (a) we see that \eqref{LL0} holds  for the eigenvalues $\bar\ll_n$ of $-L_{J,\bar V}$, i.e.
$$\bar\ll_n\ge    \ff{\dd_1}{\ll(\dd_2 n)},\ \ \ n\ge 1 $$ holds for some constants $\dd_1,\dd_2>0$.
Noting that $  \bar\ll_n=\ll_n+\ll_1\le 2\ll_n$, we prove \eqref{LL0} for   $2\dd_1 $ replacing $\dd_1$.

 \end{proof}

\section{Proof of Theorem \ref{T1.3}}

We first present a general result on the lower bound estimate by using \cite[Theorem 2.4]{W02}. Let $E,\mu, (L_0,\D(L_0)), P_t^0, (L_V,\D(L_V))$ and $P_t^V$ be in the beginning of Section 2. Assume that $P_t^0$ and $P_t^V$ are realized by a Markov process $X_t^\cdot$; i.e. for any $x\in E$, $X_t^x$ is a Markov process on $E$ starting at $x$ such that
\beq\label{FK}  P_t^0 f(x) =\E f(X_t^x),\ \ P_t^vf(x)= \E \Big[f(X_t^x) \e^{-\int_0^t V(X_s^x)\d s}\Big]\end{equation}
holds for $x\in E, t\ge 0$ and $f\in  \B_b(E)\cap L^2(\mu).$ We have the following result.

\beg{prp}\label{P4.1} Assume that $$\rr_1(t):= \|P_t^0\|_{L^1(\mu)\to L^\infty(\mu)}<\infty,\ \
 \rr_2(t):= \int_E\e^{-2t V}\d\mu <\infty,\ \ t>0.$$ Then $-L_V$ has purely discrete specturm and the eigenvalues $(\ll_n^V)_{n\ge 1}$ satisfies
 $$\ll_n^V\ge \inf_{t>0}\ff 1 {2t} \log\ff{n+1}{\rr_1(t)\rr_2(t)},\ \ n\ge 1.$$\end{prp}

 \beg{proof} Let $f\in   \B_b(E)\cap  L^2(\mu)$.  By \eqref{FK}
 and the Schwarz/Jensen  inequalities, we obtain
\beg{align*} P_t^{V} f(x) &= \E\Big[f(X_t^x)\e^{-\int_0^t V(X_s^x)\d s}\Big] \le \ss{P_t^0 f^2(x) \E\e^{-2\int_0^t V(X_s^x)\d s}}\\
&\le  \ss{\rr_1(t)}   \|f\|_{L^2(\mu)} \bigg(\ff 1 t \int_0^t P_s^0\e^{-2tV}(x)\d s\bigg)^{\ff 1 2}.\end{align*}
This implies that $P_t^V$ has a density $p_t^V(x,y)$ with respect to $\mu$ and
$$p^V_{2t}(x,x)=\int_E p_t^V(x,y)^2 \mu(\d y) \le \ff{\rr_1(t) }t \int_0^t P_s^0 \e^{-2tV}(x)\d s.$$ Since $\mu$ is $P_s^0$-invariant, this implies
$$\mu(p_{2t}^V):=\int_E \rr_{2t}^V(x,x)\mu(\d x)\le \rr_1(t) \rr_2(t),\ \ t>0.$$ Therefore,   \cite[Theorem 2.4]{W02} gives
$$\ll_n^{V} \ge \inf_{t>0} \ff 1 {2t} \log \ff{n+1} {\mu(p_{2t}^V)} \ge \inf_{t>0}\ff 1 {2t} \log\ff{n+1}{\rr_1(t)\rr_2(t)},\ \ n\ge 1.$$
  \end{proof}

We will also use the following variational formula of $\ll_n$:
\beq\label{VL}\ll_n=\inf_{(u_1,\cdots, u_{n})\in \scr S_{n}} \sup_{u\in B(u_1,\cdots, u_n)}  \scr E_{J,V}(u,u),\ \ n\ge 1,\end{equation}
where $B(u_1,\cdots, u_n):=\big\{u\in {\rm span} \{u_1,\cdots, u_n\}:\ \int_{\R^d} u(x)^2\d x =1\big\},$ and $(u_1,\cdots, u_{n})\in \scr S_n$ means that
\beq\label{U}  u_i\in \D(\scr E_{J,V}),  \ \ \int_{\R^d} (u_iu_j)(x)\d x=1_{\{i=j\}},\ \ 1\le i,j\le n.\end{equation}

\beg{proof}[Proof of Theorem \ref{T1.3}(1)]       Let $p_t^{\aa,\theta}(x,y)$ be the heat kernel of the Schr\"odinger semigroup $P_t^{\aa,\theta}$ generated by
$$L_{\aa,\theta}:= -(-\DD)^{\ff\aa 2} -|\cdot|^\theta. $$ It is well known that
 $$\rr_1(t):=\sup_{x,y\in \R^d} p_t^\aa(x,y) \le c_1 t^{-\ff d \aa},\ \ t>0$$ holds for some constant $c_1>0$. Next, there exists a constant $c_2>0$ such that
 $$\rr_2(t):=\int_{\R^d} \e^{-2|x|^\theta}\d x\le c_2 t^{-\ff d \theta},\ \ t>0.$$ So, by Proposition \ref{P4.1},   the eigenvalues  $\{\ll_n^{\aa,\theta}\}_{n\ge 1}$ of $L_{\aa,\theta}$ satisfy
\beq\label{JJ2} \ll_n^{\aa,\theta}\ge \inf_{t>0}\ff 1 {2t} \log \Big[\ff{n+1}{c_1 c_2} t^{\ff{d(\aa+\theta)}{\aa\theta}}\Big]\ge c_3 n^{\ff{\aa\theta}{d(\aa+\theta)}},\ \ n\ge 1 \end{equation} for some constant $c_3>0.$

Now, let  $\EE_{\aa,\theta}$ be the Dirichlet form associated to $L_{\aa,\theta}$. By \eqref{LB} and $V(x)\ge c|x|^\theta$ for large $|x|$,
\beq\label{JJ3} \EE_{J,V}(f,f)\ge c_4\EE_{\aa,\theta}(f,f) -c_5 \int_{\R^d} f^2(x)\d x\end{equation} holds for some constants $c_4,c_5>0.$ Combining this with \eqref{VL} and \eqref{JJ2}, we prove
  \eqref{LL2} for some constant $\dd>0$ and large enough $n$. Since $\ll_n\ge \ll_1>0$, \eqref{LL2} holds for some constant $\dd>0$ and all $n\ge 1.$\end{proof}

\beg{proof}[Proof of Theorem \ref{T1.3}(2)]  Similarly to  \eqref{JJ3}, condition \eqref{UP} and $V(x)\le c|x|^\theta$ for large $|x|$ imply
$$\EE_{J,V}(f,f)\le c'\EE_{\aa,\theta}(f,f) +c' \int_{\R^d} f^2(x)\d x$$ for some constant $c'>0$. Combining this with $\eqref{VL}$,    we may and do assume that
 \beq\label{UP'}V(x)= |x|^\theta, \ \  J(x,y) =  \ff{1 }{|x-y|^{d+\aa }},\ \ x,y\in\R^d.\end{equation}

(1) To construct suitable functions  $(u_1,\cdots, u_{n})\in \scr S_n$,   let
\beg{align*} &\xi(k)= k^{\ff{\aa}{\theta+\aa}},\ \ k\ge 1,\\
&h_k(s)= \min\big\{(s-\xi(k))^+, (\xi(k+1)-s)^+\big\},\ \ s\in\R, k\ge 1.\end{align*} Obviously, there exists a constant $c_0>1$ such that
\beq\label{G0} c_0^{-1} k^{-\ff{\theta}{\theta+\aa}}\le \xi(k+1)-\xi(k)\le c_0k^{-\ff{\theta}{\theta+\aa}},\ \ k\ge 1.\end{equation}
Next, for any $n\ge 1$, let
$$G_n=\{1,2,\cdots, n\}^d=\big\{(k_1,\cdots, k_d):\ 1\le k_i\le n, 1\le i\le d\big\}.$$ Define
$$u_\k(x):=\prod_{i=1}^d h_{k_i}(x_i),\ \ \ \k:=(k_1,\cdots, k_d)\in G_n.$$
Then
$$\int_{\R^d} (u_\k u_{\k'})(x)\d x=0,\ \ \k\ne \k',$$
and
\beq\label{G*}\beg{split} &\int_{\R^d} u_\k(x)^2 \d x =\prod_{i=1}^d \bigg(2\int_{\xi(k_i)}^{\ff 1 2 (\xi(k_i)+ \xi(k_i+1))} (s-\xi(k_i))^2\d s\bigg)\\
&=\prod_{i=1}^d \bigg(2\int_{0}^{\ff 1 2 (\xi(k_i+1)-\xi(k_i))} s^2\d s\bigg)= \ff 1 {6^d}  \prod_{i=1}^d \big( \xi(k_i+1)- \xi(k_i)\big)^3.\end{split} \end{equation}So, there exists a constant $c_1,c_2>0$ such that
\beq\label{G1} c_1  \prod_{i=1}^d k_i^{-\ff{3\theta}{\theta+\aa}}\le I_\k:= \int_{\R^d}u_\k(x)^2\d x\le c_2 \prod_{i=1}^d k_i^{-\ff{3\theta}{\theta+\aa}},\ \ \k\in G_n, n\ge 1.\end{equation}
Since every $u_\k$ is Lipschitz continuous with compact support, we have $u_\k\in \D(\EE_{J,V})$. Since $G_n$ contains $n^d$ many numbers, we have  $\big( I_\k^{-\ff 1 2}u_\k: \k\in G_n)\in \scr S_{n^d}$ so that by \eqref{VL},
\beq\label{GG0} \ll_{n^d}\le \sup_{u\in B(u_\k:\ \k\in G_n)}  \EE_{J,V}(u,u),\ \ \  n\ge 1.\end{equation}

(2) To estimate the upper bound in \eqref{GG0}, we first  bound    $\EE_{J,V}(u_\k,u_\k)$ by $I_k$. We Observe that
\beq\label{OBS} \beg{split} &u_\k(x+z)-u_\k(x)\ne 0 \ \text{implies}\\
&\xi(k_i)-|z_i|\le x_i\le \xi(k_i+1)+|z_i|\ \text{for\ all\ } 1\le i\le d.\end{split}\end{equation} Indeed, $u_\k(x+z)-u_\k(x)\ne 0$ only if at leat one of $u_\k(x+z)$ and $u_\k(x)$ is non-zero, so that either $\xi(k_i)\le x_i\le \xi(k_i+1)$ or $\xi(k_i)\le x_i+z_i\le \xi(k_i+1)$ holds, which ensures the assertion.     For any  $z=(z_1,\cdots, z_d)\in \R^d$, we take
$$z^0=0,\ \ z^i= (z_1,z_2,\cdots, z_i, 0,\cdots, 0)\in \R^d, \ \ \ 1\le i\le d.$$
By \eqref{UP'},  we obtain
\beq\label{G2} \beg{split} &\int_{\R^d\times\R^d} \big(u_\k(x+z)-u_\k(x)\big)^2 J(x,x+z)\d z\\
&\le  d  \sum_{i=1}^d \int_{\R^d\times\R^d}  \big(u_\k(x+z^i)-u_\k(x+z^{i-1})\big)^2|z|^{-(d+\aa)} \d z,\ \ \ x\in \R^d.\end{split}\end{equation}
According to  \eqref{OBS},  $u_\k(x+z^i)-u_\k(x+z^{i-1})\ne 0$ implies
$$\xi(k_i)-|z_i|\le x_i\le \xi(k_i+1)+|z_i|.$$
This together with the Lipschitz continuity of $u_\k$ implies
\beg{align*}& 1_{\{|z_i|\le \xi(k_i+1)-\xi(k_i)\}} \big(u_\k(x+z^i)-u_\k(x+z^{i-1})\big)^2\\
&\le 1_{\{|z_i|\le \xi(k_i+1)-\xi(k_i)\}} 1_{\{\xi(k_i)-|z_i|\le x_i \le \xi(k_i+1)+|z_i|\}} z_i^2 \prod_{j\ne i} (h_{k_j}(x_j+(z^{i-1})_j))^2\\
&\le 1_{\{|z_i|\le \xi(k_i+1)-\xi(k_i)\}} 1_{\{2\xi(k_i)-\xi(k_i+1)\le x_i \le 2\xi(k_i+1)-\xi(k_i)\}} z_i^2 \prod_{j\ne i} (h_{k_j}(x_j+(z^{i-1})_j)))^2,\ \ x\in\R^d.\end{align*}
As in \eqref{G*} we have  $$\int_{\R^{d-1} } \prod_{j\ne i} (h_{k_j}(x_j+(z^{i-1})_j)))^2\prod_{j\ne i} \d x_j=6^{1-d} \prod_{j\ne i} (\xi(k_j+1)-\xi(k_j))^3.$$
Combining these with  \eqref{G0} and \eqref{G1}, we arrive at
\beq\label{G3} \beg{split}&\int_{\R^d} \d x\int_{\{z: |z_i|\le \xi(k_i+1)-\xi(k_i)\}} \big(u_\k(x+z^i)-u_\k(x+z^{i-1})\big)^2 |z|^{-(d+\aa)}   \d z\\
&\le c_3 \big\{\prod_{j\ne i} (\xi(k_j+1)-\xi(k_j))^3\big\}\cdot \big\{\xi(k_i+1)-\xi(k_i)\big\} \\
&\quad\times\int_{\{z: |z_i|\le \xi(k_i+1)-\xi(k_i)\}} z_i^2 |z|^{-(d+\aa)}   \d z\\
&\le c_4 I_\k k_i^{\ff{2\theta}{\theta+\aa}} \int_0^{\xi(k_i+1)-\xi(k_i)} s^2\d s \int_{\R^{d-1}}  (|s+|\tt z|)^{-(d+\aa)} \d\tt z\\
&\le c_5 I_\k k_i^{\ff{2\theta}{\theta+\aa}} \int_0^{\xi(k_i+1)-\xi(k_i)}   \ff{s^2}{s^{1+\aa}} \d s \\
&= c_5 I_\k k_i^{\ff{2\theta}{\theta+\aa}}  \ff {(\xi(k_i+1)-\xi(k_i))^{2-\aa}}{2-\aa} \\
&\le c_6 I_\k k_i^{\ff{2\theta}{\theta+\aa}}k_i^{-\ff{\theta(2-\aa)}{\theta+\aa}}= c_6 I_\k k_i^{\ff{\theta\aa}{\theta+\aa}},\ \ k\in G_n, 1\le i\le d, n\ge 1\end{split}\end{equation}
for some constants $c_3,c_4,c_5, c_6>0$.  On the other hand, by \eqref{G0},
\beg{align*} &\int_{\R^d} \d x \int_{\{z: |z_i|> \xi(k_i+1)-\xi(k_i)\}} \big(u_\k(x+z^i)-u_\k(x+z^{i-1})\big)^2 |z|^{-(d+\aa)}   \d z\\
&\le 2 \int_{\{z: |z_i|> \xi(k_i+1)-\xi(k_i)\}}  |z|^{-(d+\aa)} \d z \int_{\R^d}   \big(u_\k(x+z^i)^2+u_\k(x+z^{i-1})^2\big) \d x \\
&= 4 I_\k  \int_{\{z: |z_i|> \xi(k_i+1)-\xi(k_i)\}}  |z|^{-(d+\aa)}   \d z\\
&\le 4 I_\k  \int_{\{ |z|> \xi(k_i+1)-\xi(k_i)\}}  |z|^{-(d+\aa)}   \d z\\
&\le c_7 I_\k  \big(\xi(k_i+1)-\xi(k_i)\big)^{-\aa} \\
&\le c_8 I_\k k_i^{\ff{\theta\aa}{\theta+\aa}},\ \  \k\in G_n, 1\le i\le d, n\ge 1\end{align*}
holds for some constants $c_7,c_8>0$. Combining this with \eqref{G2}, \eqref{G3}, and  that $k_i\le n$ for $\k\in G_n$, we arrive at
\beq\label{*J0}  \EE_J(u_\k,u_\k):= \int_{\R^d\times\R^d}  \big(u_\k(x+z)-u_\k(x)\big)^2J(x,x+z) \d x \d z \le c_9 I_\k n^{\ff{\theta\aa}{\theta+\aa}},\ \ \k\in G_n, n\ge 1 \end{equation}
for some constants $c,c_9>0$.  Finally, by \eqref{UP'} and noting that ${\rm supp}\,u_\k\subset \{|\cdot|\le d \xi(n+1)\}$  for $\k\in G_n$,  we have
\beq\label{*J1}\int_{\R^d} (u_\k^2V)(x)\d x\le c_{10} I_\k\xi(n+1)^\theta \le c_{11}I_\k n^{\ff{\theta\aa}{\theta+\aa}}, \ \ \k\in G_n, n\ge 1\end{equation} for some constants $c_{10}, c_{11}>0$.

(3) Now, for any $u\in B(u_\k:\ \k\in G_n)$, we have
$$u= \sum_{\k\in G_n} a_k \ff{u_\k}{\ss{I_k}},\ \ \sum_{\k\in G_n} a_k^2=1.$$
Since $u_\k u_{\k'}=0$ for $\k\ne \k'$, \eqref{*J1} implies
\beq\label{*J2} \int_{\R^d} (u^2V)(x)\d x =\sum_{\k\in G_n} a_k^2 \int_{\R^d}(u_\k^2V)(x)\d x \le c_{11} n^{\ff{\theta\aa}{\theta+\aa}}.\end{equation}
On the other hand, by \eqref{OBS},   $u_\k(x+z)-u_\k(x)\ne 0$ implies
$$\xi(k_i)-|z_i|\le x_i\le \xi(k_i+1)+|z_i|,\ \ 1\le i\le d.$$ So, for any $\k,\k'\in G_n$,
\beq\label{YQ} |u_\k(x+z)-u_\k(x)|\cdot |u_{\k'}(x+z)-u_{\k'}(x)|\ne 0\end{equation}  only if
$$|z_i|\ge \ff 1 2 \big\{\xi(k_i\lor k_i')-\xi(k_i\land k_i' +1)\big\},\ \ 1\le i\le d.$$ Noting that
$|k_i-k_i'|\ge 2$ implies
\beg{align*}&\ff 1 2 \big\{\xi(k_i\lor k_i')-\xi(k_i\land k_i' +1)\big\}\ge   \ff 1 2 \big\{\xi(k_i\lor k_i')-\xi(k_i\lor k_i' -1)\big\}\\
&\ge c_{12} (k_i\lor k_i')^{-\ff{\theta}{\theta+\aa}}\ge c_{12} n^{-\ff{\theta} {\theta+\aa}}\end{align*}
for some constants $c_{12}>0$. Then, when  $\|\k-\k'\|_\infty :=\max_{1\le i\le d} |k_i-k_i'|\ge 2$,    \eqref{YQ} implies
$|z|\ge c_{12} n^{-\ff{\theta}{\theta+\aa}}.$ Combining this with \eqref{UP'} and \eqref{*J0}, we arrive at
\beg{align*} &\EE_J(u,u)  \le c_2 \int_{\R^d\times \R^d} 1_{\{|z|\ge c_{13} n^{-\ff{\theta}{\theta+\aa}}\}} \ff{|u(x+z)-u(x)|^2}{|z|^{d+\aa}}\d x\d z\\
&\quad + \int_{\R^d\times \R^d} 1_{\{|z|< c_{12} n^{-\ff{\theta}{\theta+\aa}}\}}  |u(x+z)-u(x)|^2J(x,x+z) \d x\d z\\
&\le 2 c_2 \int_{\R^d\times \R^d} 1_{\{|z|\ge c_{12} n^{-\ff{\theta}{\theta+\aa}}\}} \ff{u(x+z)^2+ u(x)^2}{|z|^{d+\aa}}\d x\d z\\
&\quad + \sum_{\k,\k'\in G_n, \|\k-\k'\|_\infty\le 1} |a_\k a_{\k'}| \int_{\R^d\times\R^d} |u_\k(x+z)-u_\k(x)|\cdot |u_{\k'}(x+z)-u_{\k'}(x)|J(x,x+z)\d x\d z\\
&\le 4 c_2 \int_{\{|z|\ge c_{12} n^{-\ff{\theta}{\theta+\aa}}\}} \ff{\d z}{|z|^{d+\aa}} + \ff 1 2 \sum_{\k,\k'\in G_n, \|\k-\k'\|_\infty\le 1} \bigg(\ff{a_\k^2\EE_J(u_\k,u_\k)}{I_\k} + \ff{a_{\k'}^2\EE_J(u_{\k'},u_{\k'})}{I_{\k'}} \bigg)\\
&\le c_{13} n^{\ff{\theta \aa}{\theta+\aa}}\end{align*} for some constant $c_{13}>0$, where we have used the fact that
$$\int_{\R^d} u(x)^2 \d x= \int_{\R^d} u(x+z)^2\d x=1.$$
This together with \eqref{*J2} yields
$$\EE_{J,V}(u,u)\le c  n^{\ff{\theta \aa}{\theta+\aa}},\ \ n\ge 1, u\in B(u_\k:\ \k\in G_n)$$ for  some constant $c>0$.
Therefore, by \eqref{GG0} we obtain
 $$\ll_{n^d}\le c n^{\ff{\theta\aa}{\theta+\aa}},\ \ \ n\ge 1.$$
For any $n\ge 1$, letting $r_n=\inf\{i\in \Z_+: i\ge n^{\ff 1 d}\},$ we arrive at
$$\ll_n\le \ll_{(1+r_n)^d}\le c (1+r_n)^{\ff{\theta\aa}{\theta+\aa}}\le C n^{\ff{\theta\aa}{d(\theta+\aa)}}$$ for some constant $C>0$.
\end{proof}

\paragraph{Acknowledgement.} The authors would like to thank Professors Mateusz Kwa\'snicki,  Jian Wang and the referee for helpful comments.
In particular, the sharp lower bound in Theorem \ref{T1.3} is observed by Jian Wang.

\end{document}